\documentstyle[aps,epsf,preprint]{revtex}

\newcommand\beq{\begin{equation}}
\newcommand\eeq{\end{equation}}
\newcommand\bea{\begin{eqnarray}}
\newcommand\eea{\end{eqnarray}}

\begin{document}
\draft
%\preprint
%\widetext

\title{A new method for numerical inversion of the 
         Laplace transform. }
\author{Bruno H\"upper and Eli Pollak} 
\address{Chemical Physics Department, Weizmann Institute of Science, 76100 Rehovot, Israel}
 
\maketitle

\begin{abstract}
A formula of Doetsch ({\em Math.\ Zeitschr.} {\bf 42}, 263 (1937)) 
is generalized and used to numerically invert the one-sided Laplace 
transform ${\hat C}(\beta)$. The necessary 
input is only the values of ${\hat C}(\beta)$ on the positive 
real axis. The method is applicable provided that the functions $\hat{C}(\beta)$ 
belong to the function space 
$L^2_\alpha$ defined by the condition that
 $ G(x) = e^{x\alpha}\hat{C}(e^x),~ \alpha > 0$ has to be square integrable. 
This space includes sums of exponential decays 
${\hat C}(\beta)=\sum_n^{\infty}a_n e^{-\beta E_n}$,
e.g. partition functions with $a_n = 1$.

In practice, the inversion algorithm consists of two subsequent 
fast Fourier transforms. High accuracy inverted data can be obtained, 
provided that the signal is also highly accurate. 
The method is demonstrated for a harmonic 
partition function and resonant transmission through a barrier. 
We find accurately inverted functions even in the presence of noise.

\end{abstract}

\newpage

\renewcommand{\theequation}{1.\arabic{equation}}
\setcounter{equation}{0}
\setcounter{section}{0}
\section{Introduction}

It is often relatively easy to compute the Laplace transform 
\begin{eqnarray}
{\hat C}(\beta)\equiv\int_0^{\infty}e^{-\beta E}{\cal{C}}(E)dE
\label{1.1}
\end{eqnarray}
of a function rather than the function itself. Similarly, it is often known 
how to compute a function on the imaginary axis and it is desirable to have 
a useful method for analytic continuation of the function to real time. 
Perhaps the most notable example is the computation of the propagator
$<x\vert e^{-itH/\hbar}\vert x'>$ which is very difficult because of the 
sign problem but which is  straightforward in imaginary time 
$t=-i\hbar\beta$. A 'good' Laplace inversion methodology would 
solve both of these issues. 
The difficulty is that the inverse Laplace transform is 
known to be an ill-posed problem, since the addition of a small perturbation 
(for example $(\beta-1-ib)^{-1}$) to the image $\hat{C}(\beta)$ 
leads to a non-vanishing contribution
(i.e. $\exp\{(1 + ib)E\}$) even in the limit of a very small perturbation 
(large $b$) \cite{Cope:1990}.

Different numerical methods have been worked out to attempt at overcoming 
this problem \cite{Davies:1979},
\cite{Jarrell:1996}. They divide roughly into five classes: 
The Fourier methods \cite{Dubner:1968} which  
discretize the Bromwich inversion formula \cite{Doetsch:1943}
\begin{eqnarray}
    {\cal{C}}(E) = \frac{1}{2\pi} \int_{\sigma - i\infty}^{\sigma + i\infty}
     e^{\beta E} \hat{C}(\beta) d\beta.
\label{1.2}
\end{eqnarray}
This requires  knowledge of  the function in the complex plane and 
so does not really solve the problem.

The next two classes are based on the idea that the original function
${\cal{C}}(E)$ may be expanded
into a basis set of functions whose transforms are known. 
To this category belong the linear least-squares fitting methods,
where different basis sets are used, e.g. orthogonal polynomials 
\cite{Cope:1990,Piessens:1972}, sums of exponentials 
\cite{Papoulis:1956,Provencher:1975},
rational fractions or continued fractions \cite{Schlessinger:1967},
as well as others \cite{Schuettler:1986}.
Nonlinear least-square fits are necessary if the signal 
is decomposed directly into a sum of exponentials with unknown
decay rates $E_n$ and coefficients $a_n$ \cite{Rice:1969,Braess:1967}.
With both methods, it is difficult to treat signals accurately
which are of the form $\hat{C}(\beta) = \sum_{n=1}^\infty
a_n e^{-E_n \beta}$. 
The Laplace transform of a polynomial-type basis possesses singularities
and is inadequate for a fit to exponentials. 
On the other hand, an exponential with a non-integer decay-rate cannot
correctly be approximated by exponentials of the sort $e^{-n\beta}$. 
As a result of these difficulties, these methods are able to 
give at most five exponentials. For other signals, such 
as rational polynomials, they have proved to be very accurate.

Another approach is the singular value decomposition method (SVD) 
\cite{Bertero:1988,Linz:1994} which 
is based on the theory of inverse problems. 
This method transfers the inverse Laplace transform into a matrix
equation and the problem is transformed into the inversion of a nearly
singular matrix, an ill-posed problem as well \cite{Tikhonov:1977}. 

The fifth and most recent approach is the  maximum entropy method 
\cite{Jarrell:1996,Gubernatis:1991}.
In this method the entropy of the spectrum (which means in this context
the number of ways of reproducing the spectrum) 
subject to a certain constraint is maximized. 
This approach allows to incorporate
prior knowledge about the solution. Maximum entropy inversion is 
a nonlinear method, this complicates the practical application. However, 
it has proved its usefulness in recent computations, see for example 
Refs. \cite{Gallicchio:1994,Gallicchio:1996,Kim:1997}.
The last two methods, maximum entropy and SVD,
 have recently been compared in simulating the
electronic absorption spectrum of a chromophore coupled to a condensed
phase environment and it turned out that the maximum entropy method
is just able to reproduce the classical, smooth Franck-Condon 
contribution to the spectrum whereas SVD is capable of resolving finer
oscillatory details \cite{Egorov:1997}.

In this paper we will resurrect an old formula, derived by 
Paley and Wiener \cite{Paley:1934} and by Doetsch \cite{Doetsch:1937}, which 
is direct, there is no need to use a basis set and there is no need to solve 
a set of nonlinear equations. The Paley and Wiener form was 
rederived by Gardner et al. \cite{Gardner:1959} 40 years ago and applied 
with limited success (due in part to computational limitations) to the 
analysis of multi-exponential decay curves. 
The old formulae were derived for  functions 
${\hat C}(\beta)$ which are $L^2$ integrable and so are not directly 
useful, for example for partition functions. We will generalize them, so that the
method includes all functions which are $L_{\alpha}^2$ integrable, 
that is that the function $ {G}(x) = e^{x\alpha}\hat{C}(e^x),
~ \alpha > 0$ is $L^2$ integrable. 
We find for an exponential series that the quality 
of the inversion depends  on the magnitude of the $n$-th exponent $E_n$. 
The smaller $E_n$, the more accurate the inversion.  This enables 
to enhance the resolution of the inverted data. 

In Section II, we derive the generalized Laplace inversion formula, 
numerical properties of the formula are discussed in Section III. 
The effect of shifting  the signal is studied in Section IV. Applications 
to the harmonic oscillator partition function 
and a model resonant transmission probability are given in Section V.  
We end with a discussion of the merits of the method 
and outline some future extensions and applications.

\renewcommand{\theequation}{2.\arabic{equation}}
\setcounter{equation}{0}
\setcounter{section}{1}
\section{The continuous set ${\cal{L}}^{-1}_\alpha$ of Inverse Laplace Transforms}

In this Section we derive and generalize a Laplace inversion formula which 
uses only the  values of the Laplace transformed function ${\hat C}(\beta)$ on 
the positive, real $\beta$ axis. 
The starting point is the one-sided Laplace integral Eq. (\ref{1.1})
for which we perform a transformation of variables
\beq
        E = e^\xi~~~~~~~~~\beta = e^x   \,.
\label{2.1}
\eeq
The motivation for this transformation, which goes back to Doetsch in 1936
\cite{Doetsch:1936}, is to substitute the Laplace kernel $e^{-\beta E}$
in which we have the product of the two variables by a different one 
which contains the sum of the two variables. As a result, 
the Laplace integral takes the form of 
a convolution. If on both sides of the Laplace integral transform
the Fourier transform is applied, a certain convolution theorem can be used
in order to express the right hand side of the integral equation as a 
product of two terms. Finally, an algebraic 
manipulation leads to the desired inversion formula. 

If we follow this route 
both sides of Eq. (\ref{1.1}) are multiplied with an exponential $e^{x\alpha}$
with $\alpha > 0$ so that:
\beq
   e^{x\alpha}\hat{C}(e^x) = \int_{-\infty}^\infty e^{\alpha(x + \xi)}
   e^{-e^{x + \xi}} \left[ e^{\xi(1 - \alpha)} {\cal{C}}(e^\xi)\right] d\xi
\label{2.2} 
\eeq
Now, the integrand on the right hand side consists 
of one part which depends only on the linear combination $x + \xi$ and a second,
braced part which depends only on $\xi$. Next, both sides 
of the equation are Fourier transformed (with respect to $x$) and an application
of the convolution theorem (which is equivalent to replacing the variable 
$x$ by $z = x + \xi$) gives 
\beq
  \int_{-\infty}^\infty ~dx e^{ixy} e^{x\alpha}\hat{C}(e^x) =
  \int_{-\infty}^\infty d\xi\left[ e^{\xi(1 - \alpha)}
   {\cal{C}}(e^\xi)\right] e^{-i\xi y}
  \int_{-\infty}^\infty dz ~e^{-e^z} e^{\alpha z} e^{izy} \,.
\label{2.3}
\eeq
The last integral can be written as
\beq  \int_{-\infty}^\infty~ dz e^{-e^z} e^{\alpha z} e^{izy} = 
      \int_0^\infty dt~ t^{\alpha + iy - 1 } e^{-t} =
  \Gamma(\alpha + iy) \,,
\label{2.4}
\eeq
where $\Gamma(x)$ denotes the Gamma function \cite{Abramowitz:1984}. 
Now, rearranging Eq. \ref{2.2} leads to:
\beq 
  \int_{-\infty}^\infty d\xi\left[ e^{\xi(1 - \alpha)}
   {\cal{C}}(e^\xi)\right] e^{-i\xi y} = \frac{1}{\Gamma(\alpha + iy)}
   \int_{-\infty}^\infty dx ~e^{ixy} e^{x\alpha}\hat{C}(e^x) \,,
\label{2.5}.
\eeq
Fourier transformation of both sides of Eq. \ref{2.5} yields the 
inversion formula
\beq
  {\cal{C}}(E = e^\xi) = \frac{e^{\xi(\alpha - 1)}}{2\pi} 
  \lim_{a \to \infty} \int_{-a}^a dy
  \frac{e^{i\xi y}} {\Gamma(\alpha + iy)}
  \int_{-\infty}^\infty dx ~e^{ixy} e^{x\alpha}\hat{C}(\beta = e^x)  \,.
\label{2.6}
\eeq

Note that the inner integral in the inversion formula 
\beq g(y) = 
  \int_{-\infty}^\infty dx~ e^{ixy} e^{x\alpha}\hat{C}(e^x)
\label{2.7}
\eeq
has the symmetry property
\beq
  g(-y) = \overline{g(y)} \,,
\label{2.8}
\eeq
and the Gamma function obeys
\beq
     \Gamma(\overline{z}) = \overline{\Gamma(z)} \,,
\label{2.9}
\eeq
(where the bar denotes complex conjugation). 
This allows us to rewrite the inversion formula in a compact form as:
\bea
  {\cal{C}}(E = e^\xi) & = & \frac{e^{\xi(\alpha - 1)}}{2\pi}\lim_{a \to \infty}
  \int_0^a dy \left( \frac{e^{i\xi y}} {\Gamma(\alpha + iy)} g(y) + 
  \frac{e^{-i\xi y}} {\Gamma(\alpha - iy)} g(-y)  \right) 
  \nonumber \\
 & = &  \frac{e^{\xi(\alpha - 1)}}{\pi} {\rm Re}~\lim_{a \to \infty}
  \int_0^a dy \frac{e^{i\xi y}} {\Gamma(\alpha + iy)} 
  \int_{-\infty}^\infty dx~ e^{ixy} e^{x\alpha}\hat{C}(e^x)
  \label{2.10}
\eea

We have to require that $e^{x\alpha}\hat{C}(e^x)$ is square integrable,
lest we encounter divergent integrals. This is not a very stringent 
requirement, as we  can vary the parameter $\alpha$ to assure convergence. 
For example, the partition function of the harmonic oscillator
\beq  
   Z(\beta) = \frac{1}{2\sinh{\beta/2}}
\label{2.11}
\eeq
leads to an $L^2$ integrable  function provided that $\alpha > 1$.
  
Historically, Eq. \ref{2.10} was first derived for $\alpha = 1$ by Paley and
Wiener \cite{Paley:1934} and then more rigorously by Doetsch 
\cite{Doetsch:1937} for the special case $\alpha = 1/2$. 
In its form with $\alpha = 1$ it was applied 40 years ago 
to the analysis of multicomponent exponential decay curves
\cite{Gardner:1959}. With the choice
$\alpha > 1$ the inversion formula is now amenable to a wider class of 
functions. Not less important is the fact that a proper choice of this 
exponent improves the numerical
performance of the integrations by an order of magnitude as  discussed in 
Appendix A.

A generalization leading to the multi-dimensional inverse Laplace transform
is straightforward, as the main steps in the derivation of formula Eq.
(\ref{2.10}) are based on the Fourier integral.
The scalar variables $E, \beta$ are replaced by $N$-dimensional vectors
${\underline E}, {\underline \beta}$ to arrive at the 
$N$-dimensional inversion formula:
\beq
      {\cal {C}}({\underline E}) = 2\frac{e^{{\underline \xi}\cdot
      ({\underline \alpha} - {\bf 1})}} {(2\pi)^N} {\rm Re} 
      \lim_{{\underline a}\to \infty} \int_0^{\underline a} d{\underline y}
         \frac{e^{i{\underline \xi}\cdot{\underline y}}}{\prod_{n = 1}^N 
         \Gamma(\alpha_n + iy_n)} \int_{-\infty}^\infty d{\underline x}~
   e^{i{\underline x}\cdot{\underline y}} e^{{\underline \alpha}\cdot
   {\underline x}} \hat{C}({\underline \beta})
   \label{2.12} \,,
\eeq
where the components of ${\underline E}$ are $(e^{\xi_1}, ...,e^{\xi_N})$ and
 ${\underline \beta} = (e^{x_1},...,e^{x_N})$. The components of
 $\underline \alpha$ may be chosen to be different for each degree of freedom.

\renewcommand{\theequation}{3.\arabic{equation}}
\setcounter{equation}{0}
\setcounter{section}{2}
\section{Numerical analysis}

In any numerical application two central questions arise:

\noindent
a) How does the accuracy with which ${\hat C}(\beta)$ is known,
or which is reduced due to noise, affect 
the accuracy of the inversion technique?

\noindent
b) What is the range of $\beta$ for which it is necessary to know the Laplace 
transformed function ${\hat C}(\beta)$ in order to obtain a 'good' 
representation of the function ${\cal {C}}(E)$?

\noindent
To answer the first question we will consider in some detail the properties 
of the inversion formula for a single exponential 
$\hat{C}(\beta)= e^{-E_0\beta}$. The original signal then is
${\cal{C}}(E)=\delta(E-E_0)$, where $\delta(x)$ is the 
Dirac $'\delta'$-function. 
The function $g(y)$, cf. Eq. \ref{2.7}, can be obtained analytically:
\beq  \label{exint}
  g(y) = E_0^{-\alpha - iy} \Gamma(\alpha + iy) \,.
\label{3.1}
\eeq
This is a rapidly decaying function in the variable $y$, since the asymptotic
behavior 
of the Gamma-function for fixed $\alpha$ and large $|y|$ is:
\beq
  |\Gamma(\alpha + iy)| \to \sqrt{2\pi} |y|^{\alpha - 1/2} e^{-\pi|y|/2}\,.
\label{3.2}
\eeq
The asymptotic behavior of $g(y)$ remains the same, even if the 
signal is a sum of many exponential terms.

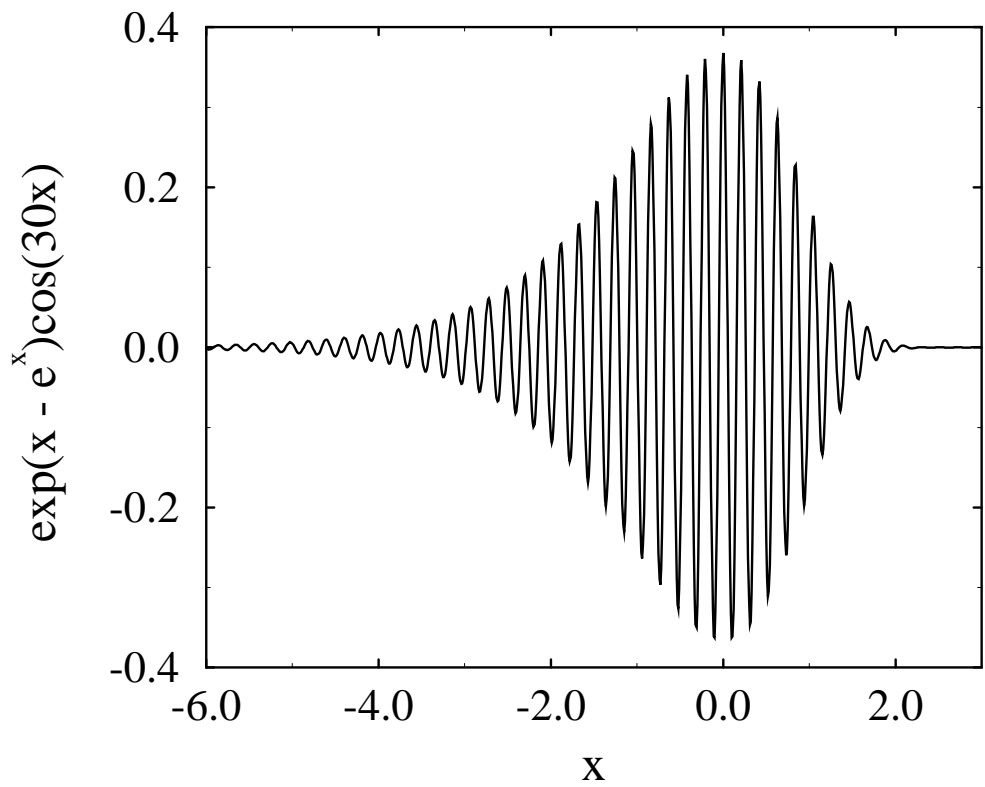
\begin{figure}
\caption[]{\label{integrandx} 
Integrand of the inverse Laplace inversion formula for a signal
of one exponential decay. The envelope is decaying exponentially
for $x \to -\infty$ and even more rapidly for $x \to +\infty$.
The rapid oscillations result in an exponentially small value 
$g(20) \approx 5\cdot 10^{-14}$ of the
integral although the integrand is of the order of unity.}
\end{figure}

As seen from the exact integration Eq. (\ref{3.1}), the Gamma-function
in the denominator of the inversion formula Eq. (\ref{2.10}) is
cancelled and the $y$-dependence is simply an oscillating function
with constant amplitude. This means that the numerically integrated 
function $g(y)$ has to be exponentially small in $y$ in order to compensate
for the exponentially increasing factor coming from the denominator. 
The integrand of $g(y)$ as shown in Fig. \ref{integrandx},
has an envelope of order unity and due to the term $e^{ixy}$,
it oscillates with increasing frequency as $y$ increases. 
In the numerical integration, it is necessary to add up many Simpson boxes
of height of the order of unity which leads to an exponentially small 
number $e^{-\pi y/2}$. The boxes have different signs, so 
for large $y$ this leads to the severe problem of adding many numbers with 
alternating signs, which differ only by a small amount. 
At this step we encounter the fact that the inverse Laplace transform
is an ill-posed problem: a small error in the signal 
\beq
  \hat{C}(\beta) = <\hat{C}(\beta)> + \delta \hat{C}(\beta) 
\label{3.3}
\eeq
leads to a large error in the integration 
\beq
   <{g}(y)> + \delta g(y) = \int K(x,y)
       (<\hat{C}(\beta)> + \delta \hat{C}(\beta)) \,.
\label{3.4}
\eeq
Let $m$ denote the number of decimal digits of the signal and assume that the 
numerical integrations are all carried out with this same accuracy. 
Then  $g(y)$ can be obtained only up to a certain maximum value 
\beq
  y_{\rm max} \approx \frac{2}{\pi}m\ln 10  
\eeq
beyond which any result is meaningless.  This implies that the 
$y$-integration has to be truncated at $y_{\rm max}$. 
For the single exponent, this leads to the result:
\beq    \label{oneexp}
  {\cal{C}}(E) = \frac{1}{\pi} E^{\alpha-1}E_0^{-\alpha} 
     \frac{\sin (y_{\rm max} \ln(E/E_0))}{\ln(E/E_0)} \,,
\label{3.5}
\eeq
which is a function peaked at $E = E_0$ with many side
oscillations. An increase of the precision $m$ allows to extend $y_{\rm max}$
resulting in a narrower peak. In the limit $y_{\rm max} \to \infty$
the exact delta function is recovered.

To appreciate this result it is instructive to compare it with an analogous 
analysis of the Fourier transform of a $\delta$-function. 
In this case, truncation of the inverse Fourier transform leads to 
the approximation: 
\beq
   {\cal{C}}(E) = \frac{1}{\pi} \frac{\sin(t_{\rm max}(E - E_0))}{E - E_0}
\label{3.6}
\eeq
If the widths $\triangle E$ of these truncated $\delta$-functions are
defined as the distance between the first two zeros on each side 
of the maximum we have for the inverse Laplace transform
\beq 
   \triangle_L E = 2E_0 \sinh\frac{\pi}{y_{\rm max}}\sim E_0\frac{2\pi}{y_{\rm max}}
\sim E_0{\pi^2\over m \ln10} \,,
\label{3.7}
\eeq
and for the inverse Fourier transform
\beq 
   \triangle_F E = \frac{2\pi}{t_{\rm max}} \,.
\label{3.8}
\eeq
Whereas for the Fourier integral the width is independent of the
position of the approximate delta, this is not true for the inverse Laplace
transform. The resolution is automatically higher for lower energy. This means 
that for a given accuracy $m$ the resolution will increase the lower
 the energy is.

Now, we come to the second task, the required range of $\beta$.

The integrand of $g(y)$ is localized in $x$: for large negative $x$, the 
exponential factor $e^{\alpha x}$ causes the integrand to decay, while 
for positive $x$, the exponentially decaying signal itself will also 
cause the integrand to decay rapidly. The decay for positive $x$ 
is  dependent on the magnitude of the lowest exponent $E_0$.
If all calculations are performed with a precision of $m$ digits, 
the value of the function is meaningful only up to the value $\beta_{\rm max}$ 
defined as:
\beq
   e^{-E_0 \beta_{\rm max}} = 10^{-m}  ~~~
    \leadsto ~~~\beta_{\rm max} = \frac{m}{E_0} \ln 10
\label{3.9}
\eeq

Since the function $g(y)$ is strongly decaying, the sampling theorem 
\cite{Shannon:1949} assures that the integrand needs to be sampled only
at relatively few points. Therefore, the numerical integration involved 
in obtaining the function $g(y)$ is straightforward.  For example, 
for $\hat{C}(\beta) = e^{-\beta}$, $\alpha = 1/2$, 
the integrand is larger than $10^{-15}$ in a range of $\triangle x = 75.0$.
Using a Simpson integration rule  with step size $h = 0.1$ leads to an 
integration error of  $\sim 10^{-16}$ for $y \in [0,20]$.
That is, we have taken only 750 sampling points for this highly accurate
integration, although the integrand varies quite rapidly for $y = 20.0$ 
with a period of $\triangle_px = 0.3141$. 

In summary, the range of $\beta$ values needed as well as the resolution with 
which the inverse transform is performed is determined by the accuracy with which 
the Laplace transformed function ${\hat C}(\beta)$ can be obtained. The lower 
the energy of the features, the larger the $\beta$ range needed but also the 
higher the resolution.  This property will be taken advantage of in the 
next section to significantly improve the algorithm.

\renewcommand{\theequation}{4.\arabic{equation}}
\setcounter{equation}{0}
\setcounter{section}{3}
\section{ Shifting the Laplace transform. }

Consider the case where the function ${\cal{C}}(E)$ is a sum of $\delta$ 
functions:
\begin{eqnarray}
{\cal{C}}(E)=\sum_{j=0}^{\infty}a_j\delta(E-E_j)
\label{4.1}
\end{eqnarray} 
where the $E_j$'s are arranged according to ascending order. The 
Laplace transformed function is
\begin{eqnarray}
{\hat{C}}(\beta)=\sum_{j=0}^{\infty}a_je^{-\beta E_j}.
\label{4.2}
\end{eqnarray}
Define the shifted Laplace transformed function as:
\begin{eqnarray}
\hat{C}(\beta,E_s)\equiv e^{\beta E_s}{\hat C}(\beta).
\label{4.3}
\end{eqnarray}
The inverse Laplace transform will lead to the function, 
\begin{eqnarray}
{\cal{C}}(E,E_s)=\sum_{j=0}^{\infty}a_j\delta[E-(E_j-E_s)] \,,
\label{4.4}
\end{eqnarray}
where all the eigenenergies have been moved closer to the origin by the 
amount $E_s$. As shown in the previous section, such a shift will lead to 
enhanced resolution in the inverted function. 

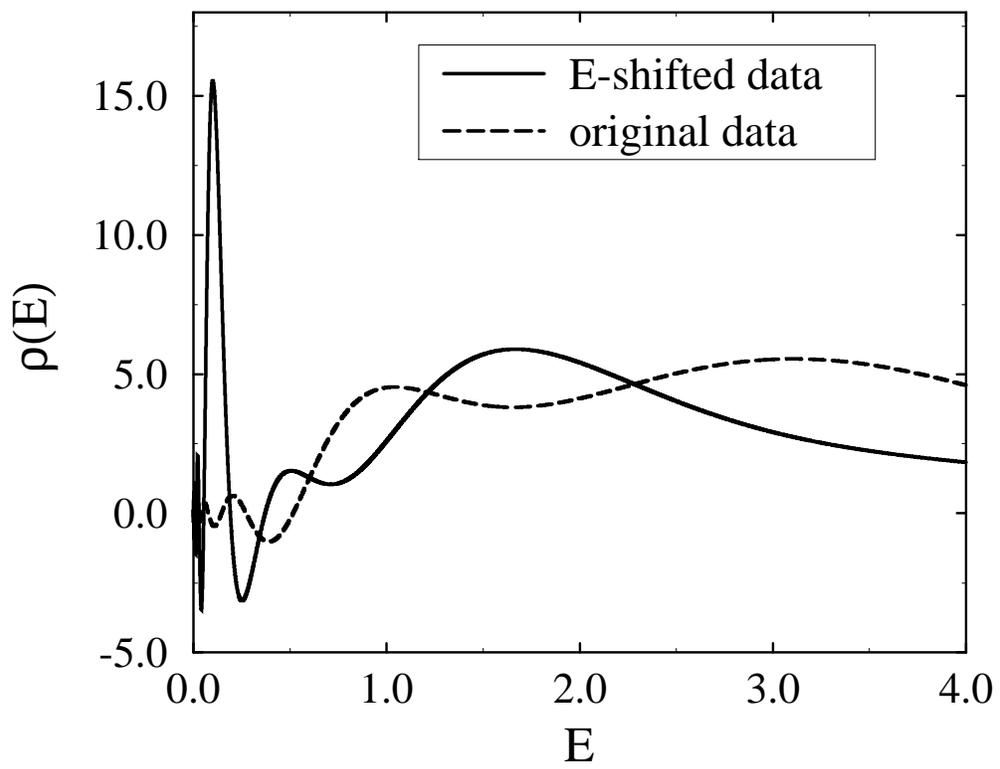
\begin{figure}
\caption[]{\label{peelshift1} 
Inverse Laplace transform of a sum of four exponential decays with decay rates
$E_n = 1,2,3,4$. The accuracy of the signal is taken as 3 decimal digits.
The inversion of the original data allows at most the estimation of the
first delta function (dashed line) at $E = 1$. The solid line shows the inversion of the data
shifted by $\delta E = 0.9$ to the left. The first maximum can now be 
estimated much more accurately.}
\end{figure}

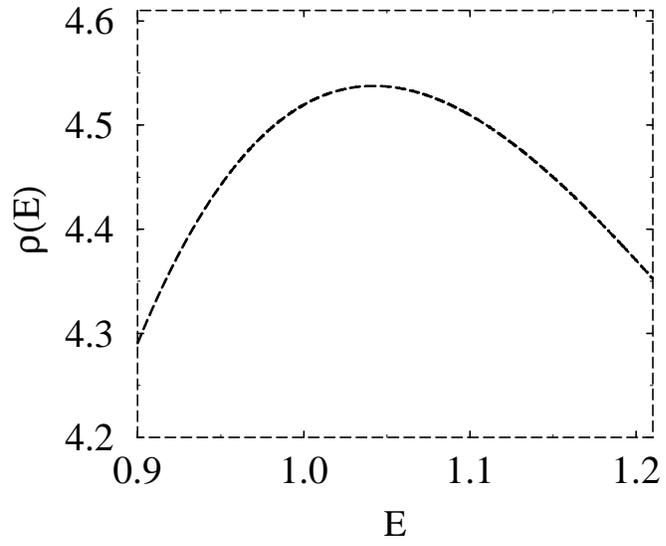
\begin{figure}
\caption[]{\label{peelshift1b} 
Magnification of the unshifted inversion of Fig. \ref{peelshift1}. The exact curve
should yield a delta function at $E = 1.0$. Due to the insufficient accuracy
of the data, the four components overlap and distort the maximum to
$E \approx 1.05$.}
\end{figure}

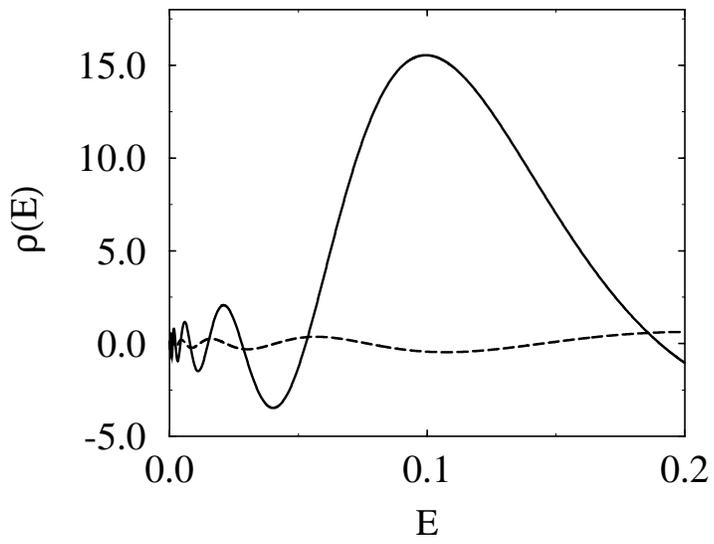
\begin{figure}
\caption[]{\label{peelshift1a} 
Magnification of the shifted inversion Fig. \ref{peelshift1}. The exact curve
should yield a delta function at $E = 0.1$. This value may now be estimated very
accurately from the shifted data, even though the accuracy ($m=3$) is low. }
\end{figure}

For illustration, let us consider four exponentials with decay rates
1,2,3,4. In Fig. \ref{peelshift1} we plot the inverse transform
with a cutoff at $y_{\rm max} = 5.0$, which means that the accuracy
of the signal is only 3 decimal digits.
Even the lowest decay rate can hardly be estimated accurately as may 
be inferred more clearly from a blow-up of the dashed line shown in 
Fig. \ref{peelshift1b}. The width of the lowest $\delta$-function is of the same 
order as the spacing and so it is hardly discernible. Shifting the function 
by $E_s = 0.9$ gives a dramatic increase in resolution. A blow-up of this 
first peak is provided in Fig. \ref{peelshift1a}. From this figure we can 
find that the maximum lies at  $E = 0.0995$. 
The price to be paid for the increased resolution is that $\beta_{max}$ 
(cf. Eq. \ref{3.9}) must be increased, since it is inversely proportional 
to the magnitude of the lowest eigenvalue which has now been reduced from 
$E_0$ to $E_0-E_s$. One may now repeat the computation, shifting the data by 
$0.999$ instead of $0.9$ and the peak will be resolved with 
even higher accuracy.  In this way, the  eigenvalue can be obtained
with arbitrary accuracy.

\renewcommand{\theequation}{5.\arabic{equation}}
\setcounter{equation}{0}
\setcounter{section}{4}
\section{Numerical Applications}
\subsection{Partition function of the harmonic oscillator}
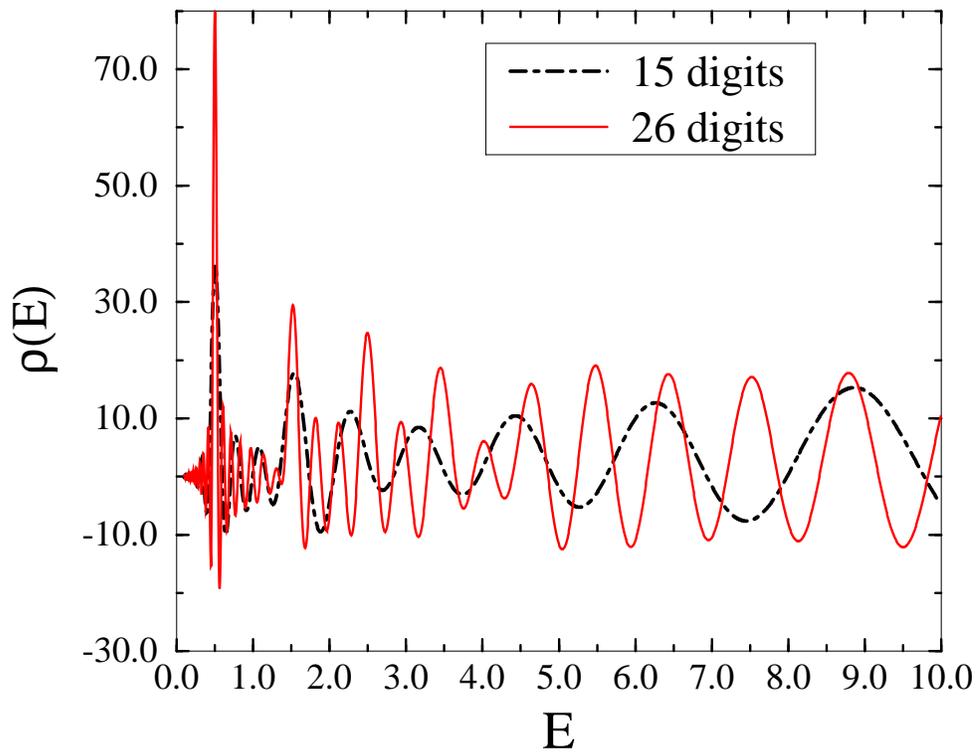
\begin{figure}
\caption[]{\label{sinhlow} 
Numerical inverse Laplace transform for the partition function of the harmonic
oscillator. The exact inverse should yield $\sum_n \delta(E - (n + 1/2))$.
The two lines correspond to different input signals whose accuracy 
(significant decimal digits) is indicated in the insert. The value of 
$\alpha=4$ was used for all computations with the harmonic oscillator 
partition function.}  
\end{figure}

The exact inversion of the partition function Eq. (\ref{2.11}) leads
to a train of delta functions at the positions of the eigenvalues of
the harmonic oscillator. This function was chosen because its
numerical Laplace inversion belongs to the most difficult class of problems.
A non-linear least-squares method (without any knowledge in advance) could
fit at most five exponentials.
On the other hand, expansions in different basis
sets converge too slowly \cite{Davies:1979}.
The inverse Laplace transform of the partition function 
was computed with different degrees
of decimal digits precision.
Fig. \ref{sinhlow} compares calculations with double precision, i.e.
15 decimal digits, and a little higher accuracy, 26 decimal digits.
Whereas for double precision only the two lowest eigenvalues can be
identified, at the  higher accuracy  the four lowest eigenvalues 
are resolved.  

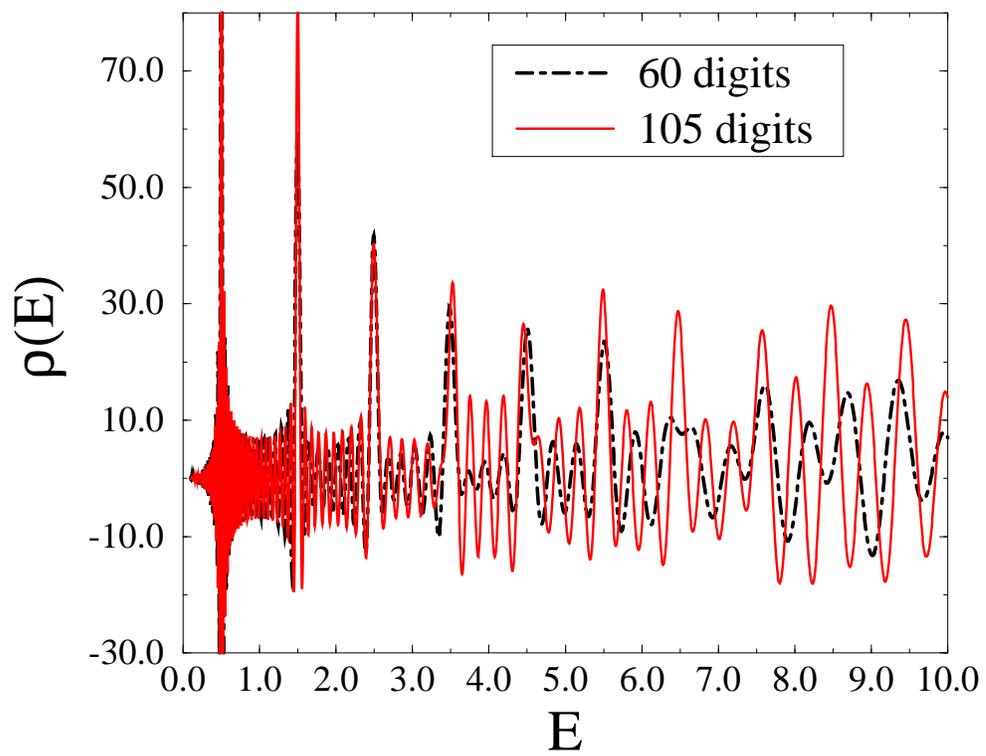
\begin{figure}
\caption[]{\label{sinhhigh} 
High precision numerical inverse Laplace transform for the partition function of the harmonic
oscillator. Other notation is as in Fig. \ref{sinhlow}. }
\end{figure}

The results of pushing the accuracy to 60 and 105 decimal digits 
precision are shown in Fig. \ref{sinhhigh}.
At 105 decimal digit precision it is possible 
to identify the eigenvalues up to the 10th level. 
The range of $\beta$ values used in all these computations
is as in Eq. (\ref{3.9}), $\beta_{\rm max} \approx 4.5 m$.
Of course, these calculations cannot be applied to data obtained from a 
Monte Carlo computation.  However, as also discussed in the next section, 
they may be used to invert basis sets which can then be 
fitted to Monte Carlo data.  These results also 
serve to demonstrate the relative simplicity and accuracy of 
the method and the fact that in principle it will work for any number of 
peaks. 

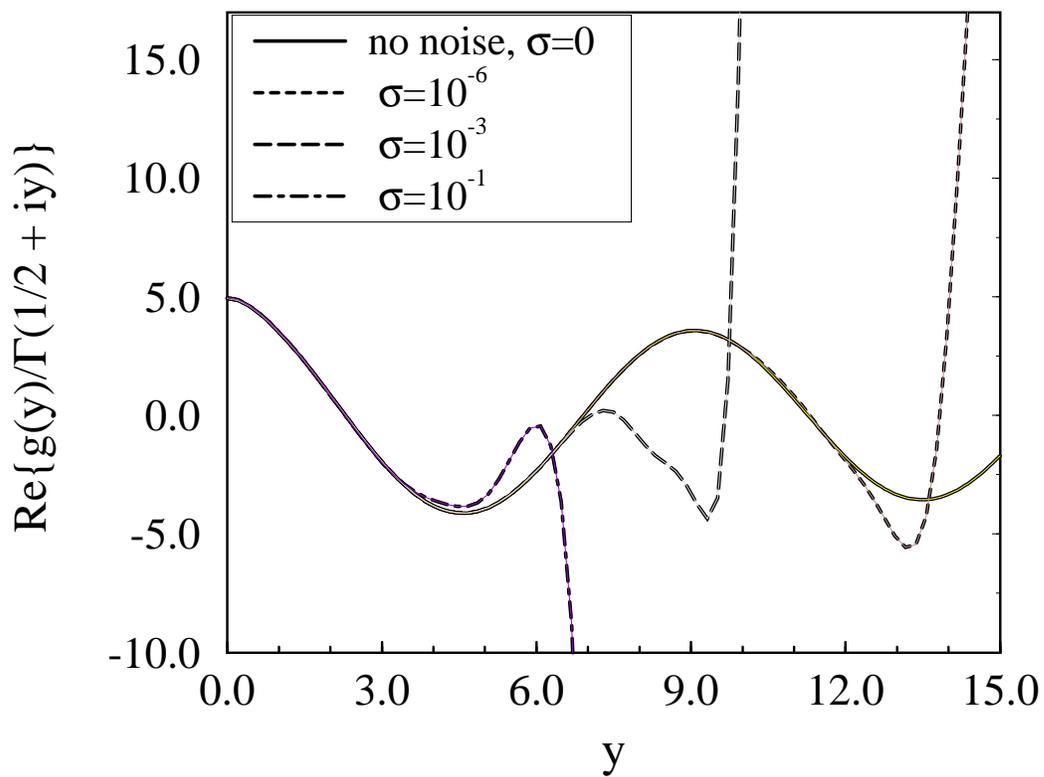
\begin{figure}
\caption[]{\label{yintegrandZ1}
Noisy data. The integrand of the real inversion formula for the 
partition function of the harmonic oscillator is plotted vs. $y$.
Gaussian noise with RMS deviation $\sigma$ as indicated is added
to the signal and this leads to a reduction of the cut-off value
for the $y$-integration.}
\end{figure}

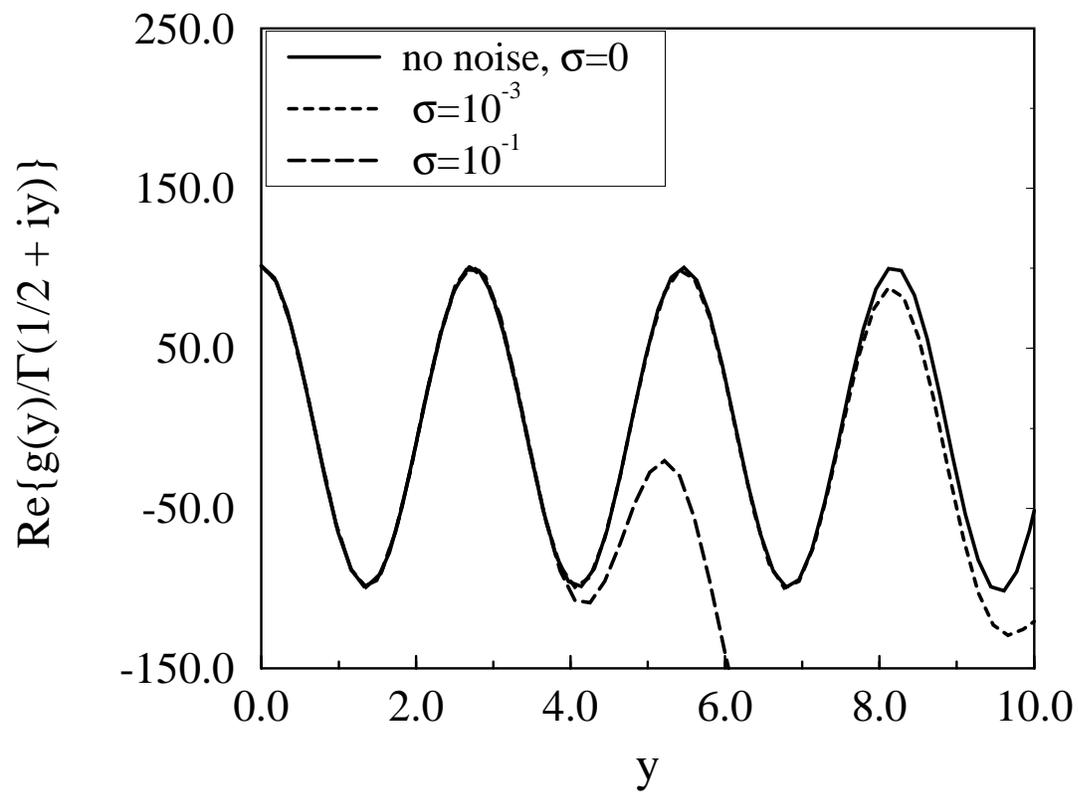
\begin{figure}
\caption[]{\label{yintegZshift} 
Noisy shifted data. 
The data used for Fig. \ref{yintegrandZ1} are shifted by 
$E_s = 0.4$ to the left. The cut-off values in $y$ remain the same, but
because of the faster oscillation of the integrand, the resolution of the final 
inversion peaks will be increased.}
\end{figure}

To test the noise-sensitivity of the inverse Laplace transform, 
we added to the signal a Gaussian distributed noise with 
zero mean and different levels of
RMS deviation $\sigma$. The signal is assumed 
to be given up to $x = 5.52$, i.e. $\beta_{\rm max} = 250$. 
Fig. \ref{yintegrandZ1} shows
 that beyond the cut-off value $y_{\rm max}$ there is an accumulation of 
 numerical errors and the signal deviates from a cosine-like wave.
This Figure also confirms that the cut-off value depends rather linearly 
on the logarithm of the RMS deviation of the noise $\sigma$. 
In Fig. \ref{yintegZshift}, the signal is shifted to the left by $E_s = 0.4$, 
so that the smallest decay rate is around $E \approx 0.1$.
The cut-off values change only slightly under the shift operation,
but the integrand contains more oscillations before the cut-off,
leading to an enhanced resolution in the peaks.

\subsection{Reflection probabilities}
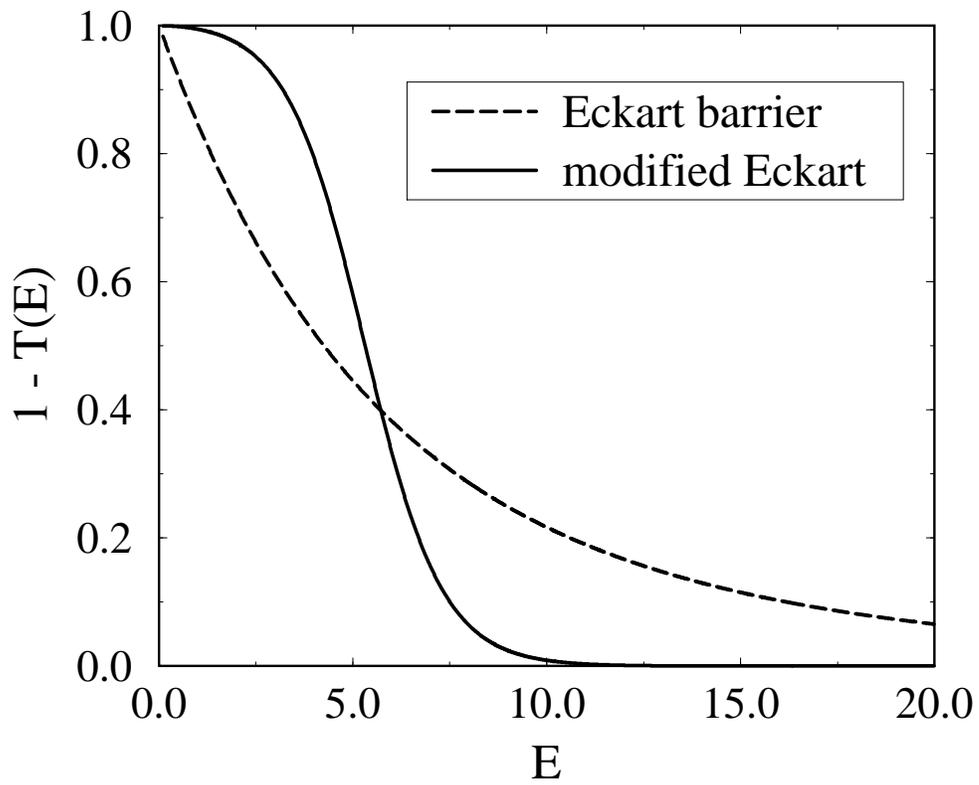
\begin{figure}
\caption[]{\label{eckart1} 
Reflection probabilities for the Eckart barrier with two different
choices of the parameters. For all reflection probabilities we used 
$\alpha=0.5$.}
\end{figure}

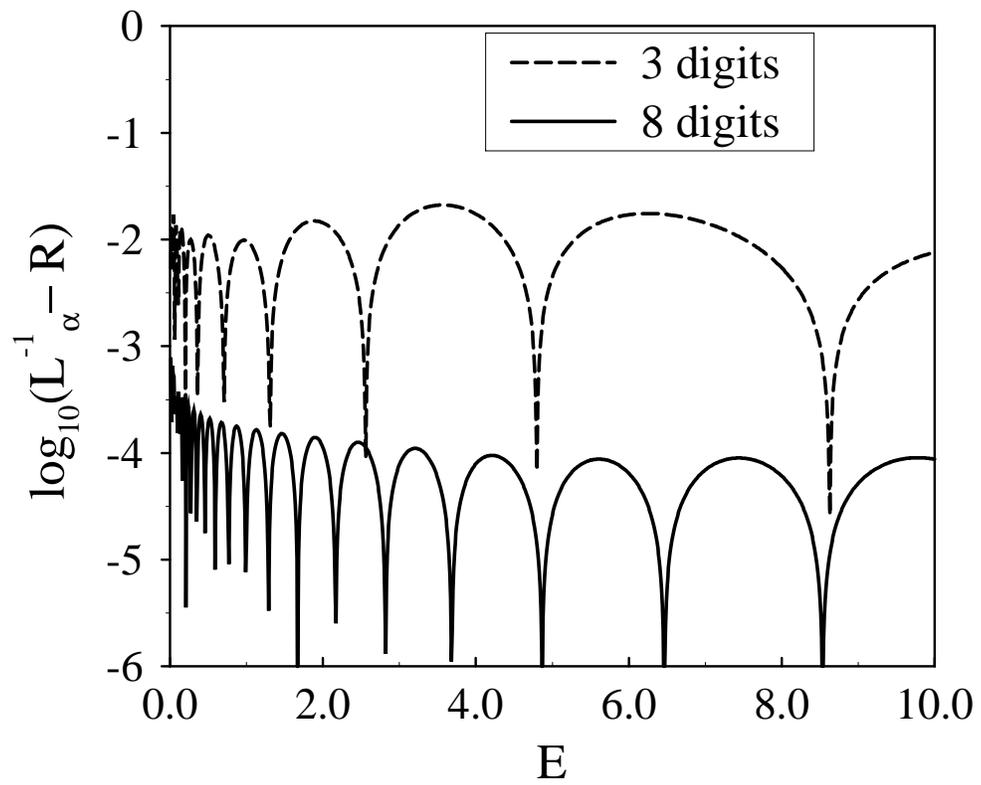
\begin{figure}
\caption[]{\label{eckarterror2}
Logarithm of the error of the inverted reflection probability of the 
Eckart potential with $\alpha = 4.0, V^{\ddag} = 5.0$. 
The signal for the inversion is obtained
by numerical Laplace transform of the exact reflection probability
and the accuracy in decimal digits 
of the numerical Laplace integral is indicated. 
The values for $y_{\rm max}$ are 5.5 and 12.0 for 3 and 8 digits accuracy 
respectively.}
\end{figure}

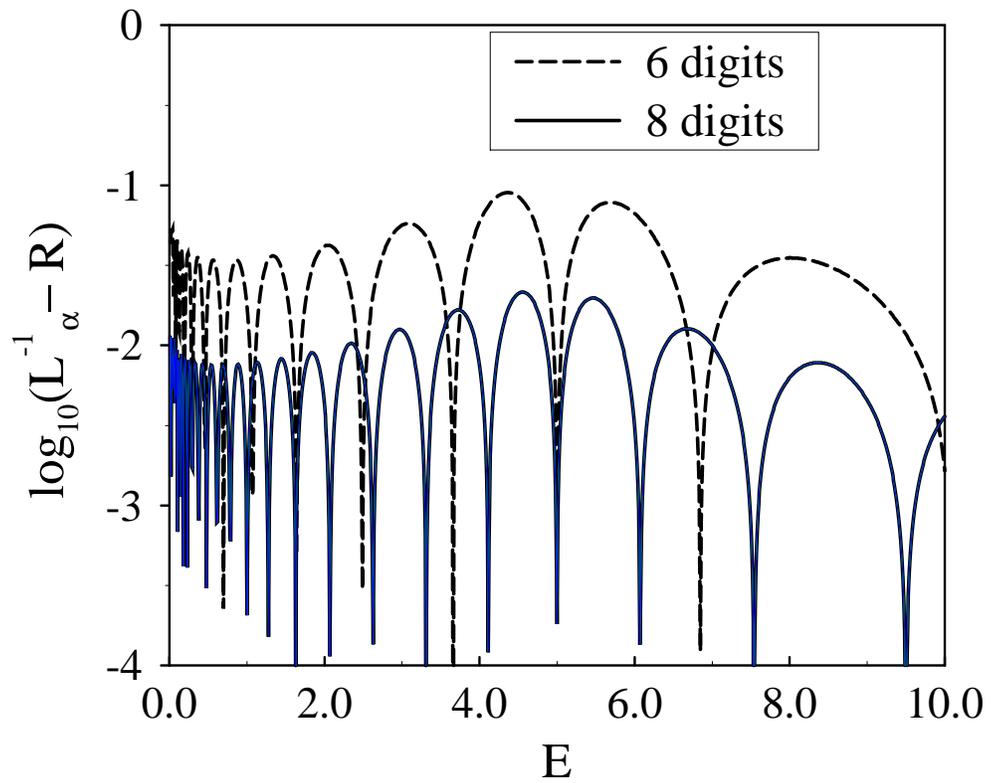
\begin{figure}
\caption[]{\label{eckarterror1}
Logarithm of the error of the inverted reflection probability of the 
Eckart potential with $\alpha = 12.0, V^{\ddag} = 5.0$. Other 
notation is as in Fig. \ref{eckarterror2}. 
The error increases near the step at $E = 5.0$ due to Gibb's phenomenon. }
\end{figure}

The Laplace transform of the reflection probability for the Eckart potential
\cite{Johnston:1961},
\beq
   R(E) = \frac{1 + \cosh(\sqrt{4\alpha^2-\pi^2})}
   {\cosh(2\alpha\sqrt{E/V^{\ddag}}) + \cosh(\sqrt{4\alpha^2-\pi^2})} \,,
\label{5.1}
\eeq
see Fig. \ref{eckart1},
is computed by numerical integration. 
Then the real inversion formula Eq. (\ref{2.10}) is used to regain the 
reflection probability. 
The difference between the exact function and the inverted one for the 
parameter choice $\alpha = 4.0, V^{\ddag} = 5.0$ is too small to be seen by 
the naked eye.  A blow-up of the error is shown in Fig. \ref{eckarterror2}. 
Even for the rather low accuracy of only 3 decimal digits the 
relative error is about $10^{-2}$, and as seen from the Figure, it 
decreases with increasing precision of the data. 
For the parameter $\alpha = 12.0, V^{\ddag} = 5.0$ the results are a 
 bit worse, as shown in Fig. \ref{eckarterror1}, 
due to the `Gibb's phenomenon` \cite{Papoulis:1962}.
Near the step, $E \approx 5.0$ the error increases significantly. 

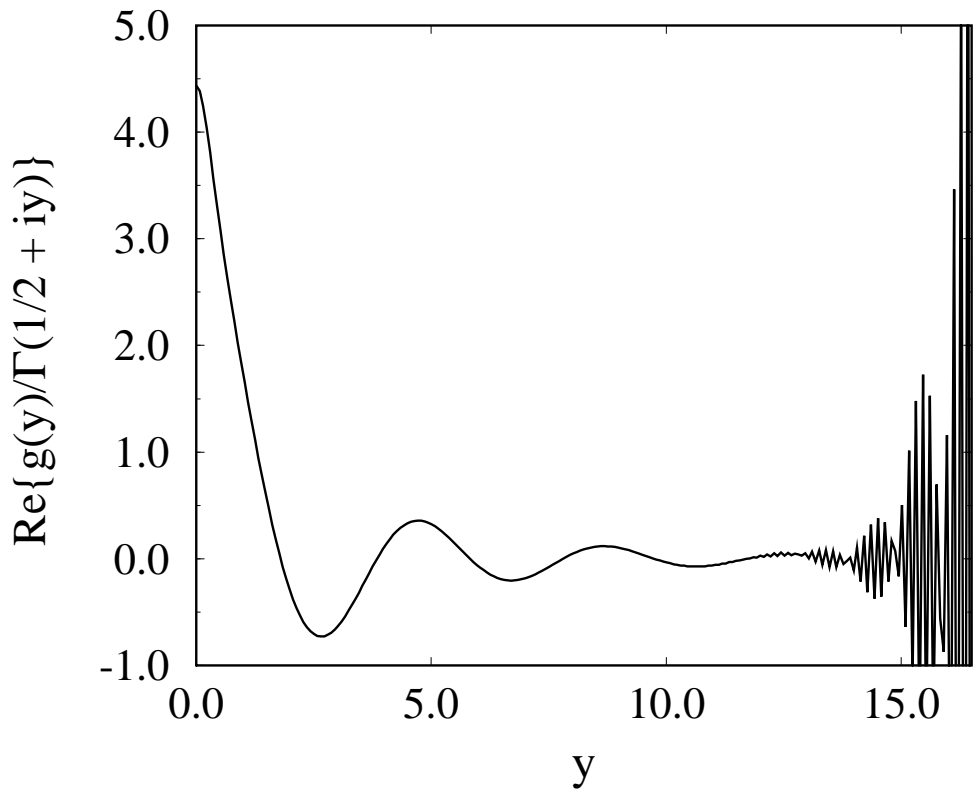
\begin{figure}
\caption[]{\label{yintegrand}
Integrand of the real inversion formula for the Eckart barrier 
reflection probability at $\alpha = 4.0, V^{\ddag} = 5.0$.
The integrand is expected to decrease like $1/(c + iy)$, but beyond
the cut-off $y_{\rm max} \approx 12.5$ artificial oscillations arise
and the integrand blows up.}
\end{figure}

In all the computations the cut-off $y_{\rm max}$ was chosen to minimize
the error: decreasing the value of $y_{\rm max}$ reduces the resolution but
increasing it leads to numerically wrong values due to
the uncertainty of the signal. In Fig. \ref{yintegrand} we show
a typical integrand ${\rm Re}\{g(y)/\Gamma(1/2 + iy)\}$: if the 
inverse Laplace transform is not known, it is easy to judge which value for
$y_{\rm max}$ has to be chosen, as the integrand decays smoothly and
then produces artificial oscillations and blows up. (For an exact step 
transmission probability $\hat{f}(\beta) = \frac{1}{\beta}e^{-V^{\ddag}\beta}$,
 the integrand goes asymptotically to 0 as $1/(\alpha - 1 + iy)$.)

\subsection{Below-barrier resonance}
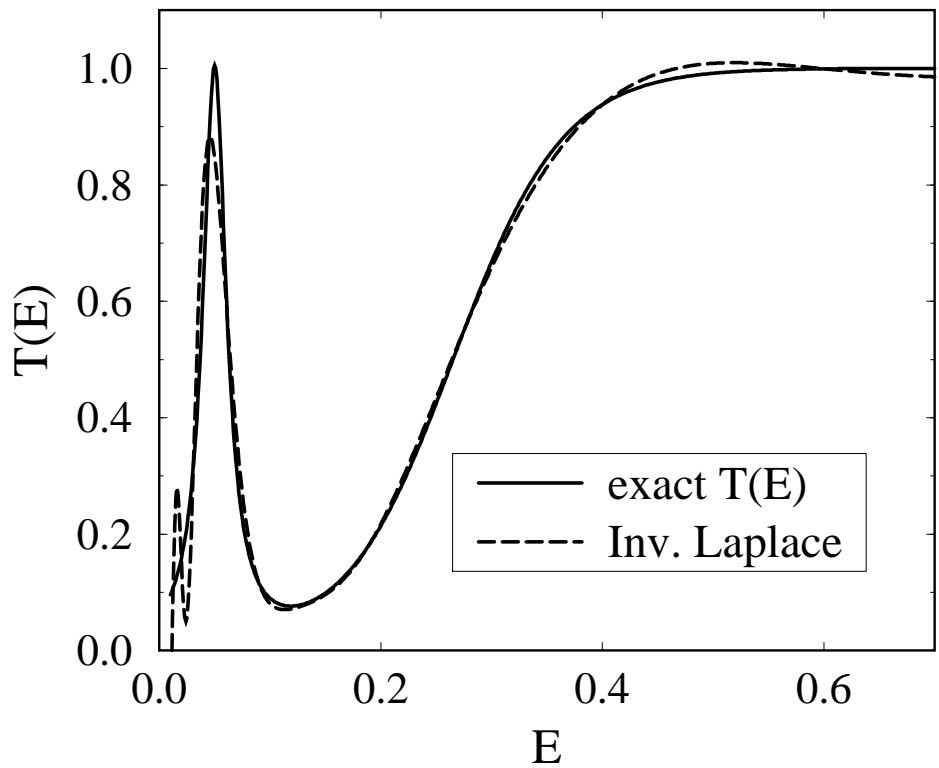
\begin{figure}
\caption[]{\label{barres1}
Numerical inverse Laplace transform for a below barrier resonance added 
to the reflection probability of the Eckart barrier 
$T_{res}(E) = \varepsilon^2/((E - E_0)^2 + \varepsilon^2)$, with 
$\varepsilon = 0.013, E_0 = 0.05$, added to the
transmission probability $T(E) = (\cosh 20E - 1)/(100 + \cosh 20E)$.
The accuracy of the data is 6 decimal digits.} 
\end{figure}

A small resonance in the form of a Lorentzian
is added to the transmission probability
\beq 
    T(E) = \frac{\varepsilon^2}{(E - E_0)^2 + \varepsilon^2} +
    \frac{\cosh E/V_0 - 1}{\cosh E/V_0 + a} \,,
\label{5.2}
\eeq
with parameters given in Fig. \ref{barres1}.
The accuracy of the data is $10^{-6}$
and the features are reproduced quite well. The oscillations
at very low energy are side oscillations of the resonance and it is 
possible to smooth them away. 
As outlined above, the resolution depends on the energy of the feature.
In order to reproduce a Lorentzian of width $\varepsilon$, it is
necessary to have at least a comparable resolution. To check whether  
the Lorentzian coming out of the inversion
is broadened because of lack of resolution, the signal can be shifted
towards lower energy. 

In this example we  took the Laplace transform of the reflection
probability with $\beta_{\rm max} = 10^6$. One may also use 
the transmission probability, however it diverges at $\beta = 0$ and 
so this requires some care.  

\renewcommand{\theequation}{6.\arabic{equation}}
\setcounter{equation}{0}
\setcounter{section}{5}
\section{Discussion}

In this paper we have resurrected and generalized a formula of Doetsch which 
enables a direct Laplace inversion of a large class of functions. By suitable 
scaling, these can include functions that are not $L^2$ integrable. Therefore 
the algorithm is directly applicable to partition functions, for example.
The method is relatively simple, all that 
is needed are two fast Fourier transforms. It is not necessary to pre-smooth 
the data. The method is controllable, the more accurate 
the Laplace inverted data, and the larger the range, the more accurate are the 
inversion results.  The parameters of the inversion are controlled by the accuracy 
of the data only.  As a result, the method is stable with respect to small 
perturbations. 

We have shown that in practice, an extremely high quality inversion can be obtained
provided that the signal is also of very high accuracy.  This is not merely an 
academic exercise.  For example, the Laguerre basis set may be taken, systematically 
inverted, and the resulting numerical functions may be stored.  Then the Laplace 
transformed function may be expanded in terms of Laguerre polynomials.  The inverted 
function is then obtained merely by reading off the inverted Laguerre functions. 
The utility of such a procedure depends on the qualitfy of the fit of the polynomials 
to the numerical Laplace transformed data.  It may be, that more sophisticated 
techniques should be used which include local smoothing of the data, such as the DAFS
methodology \cite{Hoffman:1998}. 
In any case, once the Laplace transformed data is projected onto 
standard basis sets, the high accuracy inversion may be used to obtain the inverted 
function. 

An important property of the inversion technique is the fact that the resolution 
of the resulting signal depends on the location of the signal.  The closer it is to the 
origin, the higher is the resolution.  This allows for a shifting  of 
the signal to obtain an increased resolution. The price to be paid is that 
each shift demands knowledge of the function for larger values of $\beta$. 
For analytic functions, such as the Laguerre polynomials, this does not 
present any severe difficulty, as present day computers enable computations 
with very high accuracy, which is also demonstrated for the harmonic oscillator 
partition function. 

The  Laplace inversion method presented in this paper is ideally suited for data obtained
from matrix multiplication schemes \cite{Storer:1968,Thirumalai:1983}. 
These methods produce the data at points $\beta_j = \triangle \beta 2^j$
\cite{Feynman:1965}, while  the inversion requires $\beta_j = e^{j\triangle x}$. 
 
In this paper we have not considered correlation functions. Elsewhere 
\cite{huepper:1999} we will present 
the application of the present method to spectra and correlation functions. In 
principle there is no special complication except for the fact that in some cases 
a two dimensional inverse Laplace transform has to be computed. 

We have also not considered directly the numerical analytic continuation of functions. 
As already mentioned in the Introduction, once the inverted function is obtained, 
it may be Fourier transformed to obtain the analytically continued function.  
In this sense, the inversion technique presented in this paper may be thought of as 
a representation of the complex valued Dirac $\delta$ function.  
The real question is one of practical 
usage, that is the level of  accuracy  needed to obtain the real time function from 
the imaginary time function as well as the range of $\beta$ values needed for a given 
time length. Other applications are the computation of moments of 
a probability distribution from its transform \cite{Choudhury:1996}.
These questions will be considered in future studies \cite{huepper:1999}.

\vspace*{0.5cm}

\noindent
{\bf Acknowledgements}

B. H. gratefully thanks the MINERVA foundation, Munich, for a grant and
the Weizmann Institute of Science for its hospitality. 
This work has been supported by grants from the US-Israel Binational 
Science Foundation and the Israel Science Foundation. 

\begin{appendix}

\section{Optimizing the choice of $\alpha$}
We will outline how the parameter $\alpha$ can help reduce the numerical
effort drastically, especially in high precision calculations.

The main numerical advantage of introducing $\alpha$ is a shortening of the
integration interval in $x$ needed for obtaining $g(y)$, cf. Eq. (\ref{2.7}).
The range of integration $[x_{min},x_{max}]$ 
is determined by the required accuracy $\varepsilon = 10^{-m}$.
 The negative limit is mainly fixed by the exponential $e^{\alpha x}$,
\beq
     x_{\rm min} = \frac{1}{\alpha} \ln \frac{\varepsilon}{\hat{C}(0)}\,,
\eeq
and the positive limit is due to the very rapid decay of $\hat{C}(e^x)$
which is almost independent of $\alpha$ and  
is determined by the smallest decay rate. 
The larger $\alpha$, the smaller the integration interval, but if $\alpha$ 
becomes too large the integrand increases exponentially, magnifying 
uncertainties in the signal. 

The maximum value of the integrand, if one exponential decay
$\hat{C}(\beta) = a_0e^{-E_0\beta}$ is considered, is at
$x_{\rm m} = \ln\alpha/E_0$ and the integrand $I(x)$ takes the value
\beq
   I(x_{\rm m}) = a_0 e^{\alpha(\ln \alpha/E_0 - 1)} 
       \approx   a_0 e^{\alpha\ln\alpha}   \,,
\eeq
which goes essentially as $\alpha!$. The larger $\alpha$, the more digits 
are required in the computation. 
On the other hand, the outcome of the integration
must cancel the denominator
$\Gamma(\alpha + iy)$ whose large $y$-asymptotics is given by Eq.
(\ref{3.2}). 
For a given $y_{\rm max}$ the order of magnitude of 
$\Gamma(\alpha + iy)$ divided by the integrand at $y = 0$, 
$\Gamma(\alpha) \approx \alpha^\alpha \approx e^{ \alpha\ln\alpha}$
has to be comparable to the given accuracy $\varepsilon = 
10^{-m}$:
\beq
    \frac{y_{\rm max}^{\alpha - 1/2} e^{-\pi y_{\rm max}/2}}{(\alpha-1)!} = 
          \frac{\varepsilon}{a_0}
\eeq

In summary, for large cut-off values $y_{\rm max}$ the stepsize in the $x$ integration
remains approximately the same. 
A change of $\alpha$ reduces the interval of the 
first integration, but to keep the same resolution (i.e. keep $y_{max}$ 
fixed) it is necessary to increase the precision $m$. 
We found that for $m \approx 100$, $\alpha \approx 10$ is a reasonable
choice.

\end{appendix}

\newcommand{\ACP}[1]{{\em Adv.\ Chem.\ Phys.}\/ {\bf #1}}
\newcommand{\AP}[1]{{\em Ann.\ Phys. (NY)}\/ {\bf #1}} 
\newcommand{\CMP}[1]{{\em Commun.\ Math.\ Phys.}\/ {\bf #1}} 
\newcommand{\CPL}[1]{{\em Chem.\ Phys.\ Lett.}\/ {\bf #1}} 
\newcommand{\JCP}[1]{{\em J.\ Chem.\ Phys.}\/ {\bf #1}} 
\newcommand{\JCOP}[1]{{\em J.\ Comut.\ Phys.}\/ {\bf #1}} 
\newcommand{\JETP}[1]{{\em Sov.\ Phys.\ JETP}\/ {\bf #1}} 
\newcommand{\JETPL}[1]{{\em JETP Lett.\ }\/ {\bf #1}} 
\newcommand{\JMP}[1]{{\em J.\ Math.\ Phys.}\/ {\bf #1}} 
\newcommand{\JMPA}[1]{{\em J.\ Math.\ Pure Appl.}\/ {\bf #1}} 
\newcommand{\JPA}[1]{{\em J.\ Phys. A: Math. Gen. }\/ {\bf #1}} 
\newcommand{\JPB}[1]{{\em J.\ Phys. B: At. Mol. Opt. }\/ {\bf  #1}} 
\newcommand{\JPC}[1]{{\em J.\ Phys.\ Chem.}\/ {\bf #1}} 
\newcommand{\MP}[1]{{\em Mol.\ Phys.}\/ {\bf #1}} 
\newcommand{\PLA}[1]{{\em Phys.\ Lett.}\/ {\bf A #1}} 
\newcommand{\PR}[1]{{\em Phys.\ Rev.}\/ {\bf #1}} 
\newcommand{\PRL}[1]{{\em Phys.\ Rev.\ Lett.}\/ {\bf #1}} 
\newcommand{\PRSL}[1]{{\em Proc.\ R.\ Soc.\ Lond.\ A}\/ {\bf #1}} 
\newcommand{\PRA}[1]{{\em Phys.\ Rev. A}\/ {\bf #1}} 
\newcommand{\PRB}[1]{{\em Phys.\ Rev. B}\/ {\bf #1}} 
\newcommand{\PRE}[1]{{\em Phys.\ Rev. E}\/ {\bf #1}} 
\newcommand{\PST}[1]{{\em Phys.\ Scripta }\/ {\bf A #1}} 
\newcommand{\PTPS}[1]{{\em Prog.\ Theo.\ Phys.\ Supp.}\/ {\bf #1}}
\newcommand{\RMS}[1]{{\em Russ.\ Math.\ Surv.}\/ {\bf #1}} 
\newcommand{\USSR}[1]{{\em Math.\ USSR.\ Sb.}\/ {\bf #1}} 
\newcommand{\MZ}[1]{{\em Math.\ Zeitschr.}\/ {\bf #1}} 
\renewcommand{\baselinestretch} {1}

\newpage
\begin{figure}

\vspace*{4cm}
\hspace*{-0.5cm}
\epsfbox{bild_integrandx.epsi}

\end{figure}

\vspace*{1cm}
\hspace*{6.4cm}
{\LARGE Fig. 1}

\newpage
\begin{figure}

\vspace*{4cm}
\hspace*{-0.5cm}
\epsfbox{bild_peelshift1.epsi}

\end{figure}

\vspace*{1cm}
\hspace*{6.4cm}
{\LARGE Fig. 2}

\newpage
\begin{figure}
\hspace*{2.3cm}
\epsfbox{bild_peelshift1b.epsi}
\end{figure}
\vspace*{1cm}
\hspace*{6.4cm}
{\LARGE Fig. 3}

\vspace*{1.5cm}

\begin{figure}
\hspace*{2.3cm}
\epsfbox{bild_peelshift1a.epsi}
\end{figure}

\vspace*{1cm}
\hspace*{6.4cm}
{\LARGE Fig. 4}

\newpage
\begin{figure}

\vspace*{4cm}
\hspace*{0.5cm}
\epsfbox{bild_sinhlow.epsi}

\end{figure}

\vspace*{1cm}
\hspace*{6.4cm}
{\LARGE Fig. 5}

\newpage

\begin{figure}

\vspace*{4cm}
\hspace*{-0.5cm}
\epsfbox{bild_sinhhigh.epsi}

\end{figure}

\vspace*{1cm}
\hspace*{6.4cm}
{\LARGE Fig. 6}

\newpage

\begin{figure}

\vspace*{4cm}
\hspace*{-0.5cm}
\epsfbox{bild_yintegrandZ.epsi}

\end{figure}

\vspace*{1cm}
\hspace*{6.4cm}
{\LARGE Fig. 7}
\newpage

\begin{figure}

\vspace*{4cm}
\hspace*{-0.5cm}
\epsfbox{bild_yintegZshift.epsi}

\end{figure}

\vspace*{1cm}
\hspace*{6.4cm}
{\LARGE Fig. 8}

\newpage
\begin{figure}

\vspace*{4cm}
\hspace*{0.5cm}
\epsfbox{bild_eckart1.epsi}

\end{figure}

\vspace*{1cm}
\hspace*{6.4cm}
{\LARGE Fig. 9}

\newpage
\begin{figure}

\vspace*{4cm}
\hspace*{0.5cm}
\epsfbox{bild_eckarterror2.epsi}

\end{figure}

\vspace*{1cm}
\hspace*{6.4cm}
{\LARGE Fig. 10}

\newpage
\begin{figure}

\vspace*{4cm}
\hspace*{0.5cm}
\epsfbox{bild_eckarterror1.epsi}

\end{figure}

\vspace*{1cm}
\hspace*{6.4cm}
{\LARGE Fig. 11}

\newpage
\begin{figure}

\vspace*{4cm}
\hspace*{-0.5cm}
\epsfbox{bild_yintegrand.epsi}

\end{figure}

\vspace*{1cm}
\hspace*{6.4cm}
{\LARGE Fig. 12}

\newpage
\begin{figure}

\vspace*{4cm}
\hspace*{-0.5cm}
\epsfbox{bild_barres1.epsi}

\end{figure}

\vspace*{1cm}
\hspace*{6.4cm}
{\LARGE Fig. 13}

\end{document}